\documentstyle[aps]{revtex}
\begin{document}
\def\h{{\bf h}} \def\s{{\bf s}}
\def\psic{\psi_C(\h)} \def\psif{\psi_F(\h)} \def\psia{\psi_\alpha(\h)}
\def\betac{\beta_C(\h)} \def\betaf{\beta_F(\h)} \def\betaa{\beta_\alpha(\h)}
\def\muc{\mu_C(\h)} \def\muf{\mu_F(\h)} \def\mua{\mu_\alpha(\h)}
\def\zetac{\zeta_C(\h)} \def\zetaf{\zeta_F(\h)} \def\zetaa{\zeta_\alpha(\h)}
\def\dz{\Delta\zeta}
\def\z{\zeta(\h)} \def\zo{\zeta_0(\h)} 
\def\zos{\zeta_0(\s)} \def\zosa{\zeta_0(\s+{\bf a})}
\def\e{\epsilon (c,\h)} \def\es{\epsilon_s (\h)}
\def\f{{\bf f}(\h)} \def\fs{{\bf f}(\s)}
\def\kt{k_{\rm B}T}   
\def\lang{{\cal L}}

\title{On pinning and snapping of elastic strands}
\author{Jean-Fran\c cois Palierne}
\address{ENSL-Laboratoire de Physique, URA 5672, 46 all\'ee d'Italie F-69364 Lyon Cedex 07}
\maketitle
\begin{abstract}
The transient network literature up to now has considered that the connection probability of a free strand does not depend on the strand extension, in contrast with the disconnection probability. We argue that, on thermodynamic grounds, both probabilities must have the same dependence on the strand extension. We propose a model for the pinning/snapping of elastic strands reversibly binding to sticky sites, giving explicit expressions for the connection and disconnection rates.
\end{abstract}
\section*{Introduction}
This paper addresses a seemingly paradoxical result uncovered when assessing the thermodynamic consistency of Yamamoto's transient network model (Yamamoto 1956,1957,1958, Tanaka and Edwards 1992a-d, Palierne 2000): Consider an elastic strand, i.e. a chainlike molecule or part of molecule, with both ends pinned to fixed sticky sites by reversible bounds. It is fairly obvious that the probability per unit time that a pinned end snaps back and becomes free is an increasing function of the strand extension, hence of the separation of the sticky sites. It is less obvious, however, that the probability per unit time that a free strand becomes pinned when encountering a sticky site should {\it increase in the same proportion with the strand extension}. 
This fact stems from the detailed balance condition for the connection/disconnection reaction formulated in Palierne (2000), the violation of this condition being at the root of the thermodynamic inconsistencies in Yamamoto's model and in its offspring (Palierne 2000). The argument is as follows. 

We take one end of a strand fixed at the origin of space, and consider the other end. Its position is $\h$, and we consider its pinning to a sticky site, a process hereafter called the {\it connection} of a free strand, as well as the reverse process called {\it disconnection}, whereby the pinned end snaps back and the connected strand becomes free. Connection and disconnection are instantaneous and take place without change of the connector $\h$.
(hereafter, notations close to those of Palierne 2000 will be used, except for introducing the one-strand distribution $\psia = N^{-1}\Gamma_{\alpha \h}$, where $\alpha=F$ or $C$ denotes the connection state).
In the ideal systems we consider, the strand chemical potential $\mua$ is the sum of the distribution entropy $\kt\ln\psia$ and a part $\zetaa$ depending on the number and the energy of the strand conformations compatible with the $\h$ value of the connector. The chemical potential at equilibrium is $\alpha -$ and 
$\h-$independent, $\mua=\mu^{eq}$, and the equilibrium distribution reads
\begin{equation} \label{norm}
\psia=K e^{-\zetaa/\kt}\ , \qquad
K^{-1}=\sum_{\alpha=F,C}\int e^{-\zetaa/\kt}\ d^3\bf h
\end{equation}
where $K$ ensures the normalisation $\int(\psic+\psif)d^3\h=1$. The $\h$-dependence of $\zetaa$ only involves the configuration of the strand, irrespective of its connection state, therefore the connection potential, defined as the difference
\begin{equation}
\dz=\zetac-\zetaf, 
\end{equation}
can not depend on $\h$, and $\psic$ must be proportional to $\psif$: 
$\psic=e^{-\dz /\kt}\psif$
For instance, freely-jointed strands comprising $n$ segments of length $a$ have $\zetaf=\zetac-\dz=(3\kt/2na^2){\h}^2$ and both the free and the connected strands are Gaussian:
$\psic\propto\psif\propto e^{-3{\h}^2/2na^2}$ (Palierne 2000). 

Consider now the kinetics of the connection/disconnection reaction. Let $\betac$ denote the connection rate, i.e. the probability per unit time that a free strand connects, and let accordingly $\betaf$ denote the disconnection rate. 
At equilibrium, the distributions $\psif$ and $\psic$ of free and connected strands are such that the connection/disconnection reaction rate 
\begin{equation}
v(\h) \equiv\betac\psif-\betaf\psic
\end{equation}
vanishes, resulting in the detailed balance condition
\begin{equation} 
\label{detbal}
\qquad\qquad {\betac\over \betaf}={\psic\over\psif}
=e^{-\dz /\kt} \quad \hbox{\it independent of }\h
\end{equation}
The inescapable consequence of this relation is, if we admit that the disconnection rate $\betaf$ increases with $\h$, then the connection rate $\betac$ must increase in the same proportion. In the following section, both $\betaf$ and $\betac$ will be derived from a microscopic model, and relation \ref{detbal} will be checked.

\section*{microscopic model}
We are interested in the interaction between a sticky site and one strand's end we call the sticker. The other end will therefore be kept fixed at the origin of space. Let $\{ c,\h\}$ denote the set of degrees of freedom, $\h $ being the position of the non-fixed end and $c$ denoting the other degrees of freedom. $c$ and $\h$ are assumed to vary independently and to fluctuate according to an equilibrium Gibbs distribution at temperature $T$. The energy of configuration $\{ c,\h\}$ is the sum of the strand internal energy $\e$ and the interaction energy  
\begin{equation}
\es=w\; \theta(a-|\h -\s|)
\end{equation}
describing a spherical well centered on $\s$, of depth $w$ and radius $a$, attracting the sticker and no other part of the strand ($w<0$ for an attracting well). The link between this microscopic model and macroscopic thermodynamics is provided by the quantity   
\begin{equation}
\z=-\kt \ln \sum_c e^{-{{\e+\es}\over{\kt}}}=\zo+\es
\end{equation} \begin{equation}
\hbox{ where }\ \ \zo=-\kt \ln \sum_c e^{-{{\e}\over{\kt}}} \qquad
\end{equation}
where $\sum^c$ designates teh statistical sum over all configurations keeping a given $\h$. The conector force
\begin{equation}
\f={\partial \zo \over \partial \h}
\end{equation}
is the statistical average of the microscopic force $\partial\e/\partial\h$ over the Gibbs distribution of $c$. The radius $a$ will be considered small with respect to the length scale of $\epsilon$, say $a\ll \sqrt{\epsilon/\nabla^2\epsilon}$, but the product $a\f$ need not be smaller than $\kt$. Expanding $\zo$ about $\s$ according to $\zo=\zos+\fs\cdot(\h-\s)$, we write down for further use the following integrals
\begin{equation}
I_0= \int e^{-{\zo\over \kt}} d^3\h \ \qquad\qquad\qquad\qquad\qquad\qquad 
\end{equation} 
\begin{equation}
\qquad\qquad\qquad I(\s)= \int_{|\h-\s|<a} e^{-{\zo\over \kt}} d^3\h
=4\pi a^3 e^{-{\zos\over \kt}} (u^{-2}\cosh u-u^{-3}\sinh u)
\end{equation}
\begin{equation}
J(\s)=\int_{|{\bf a}|=a} e^{-{\zosa\over \kt}} d^2{\bf a}
=4\pi a^2 e^{-{\zos\over \kt}} u^{-1}\sinh u
\end{equation}
\begin{equation}
K(\s)=\int e^{-{\zo+\es\over \kt}} d^3\h 
=I_0+(e^{-{w\over \kt}}-1)I(\s)
\end{equation}
where $u=a|\fs|/ \kt$. $I_0$ and $K(\s)$ are integrals over the whole $\h$ space, $I(\s)$ over the sphere $|\h-\s|<a$, and $J(\s)$ over the surface of the same sphere. The equilibrium Gibbs distribution of $\h$ reads
\begin{equation} \label{phi}
\phi(\h)={e^{-{\zo+\es\over \kt}}\over K(\s)}
\end{equation}
The probability $p$ of finding the sticker within the well can be written
\begin{equation}
p={e^{-{w\over \kt}}I(\s)\over K(\s)}
\end{equation}
We now introduce the kinetic properties of our model. The sticker is assumed to diffuse in a Brownian fashion both within and outside the well. The surface $|\h-\s|=a$ is assumed to be a potential barrier such that $\lambda_{\rm out}$, the probability per unit time and per unit area that the sticker crosses the barrier from the inside to the outside at point $\s+{\bf a}$ reads
\begin{equation}
\lambda_{\rm out}(\s+{\bf a})=t_{\rm out}\phi_{\rm in}(\s+{\bf a})
=t_{\rm out}\lim_{\eta\rightarrow 0^-}\phi(\s+(1+\eta){\bf a}),
\end{equation} 
$t_{\rm out}$ being the outwards transmission factor, and the probability of the reverse crossing can be written accordingly
\begin{equation}
\lambda_{\rm in}(\s+{\bf a})=t_{\rm in}\phi_{\rm out}(\s+{\bf a})
=t_{\rm in}\lim_{\eta\rightarrow 0^+}\phi(\s+(1+\eta){\bf a}),
\end{equation} 
$t_{\rm out}$ and $t_{\rm in}$ are intrinsic properties of the barrier, independent of the position on the sphere. According to eq. \ref{phi}, the limits $\phi_{\rm in}$ and $\phi_{\rm out}$ are such that
\begin{equation}
{\phi_{\rm in}(\s+{\bf a})\over \phi_{\rm out}(\s+{\bf a})} 
= e^{-{w\over \kt}}
\end{equation}
independent of the position on the sphere. The principle of detailed balance states that 
\begin{equation}
\lambda_{\rm in}=\lambda_{\rm out}
\end{equation}
at equilibrium, so the transmission factors are linked to the well depth by the customary relation
\begin{equation}
{t_{\rm in}\over t_{\rm out}}= e^{-{w\over \kt}}
\end{equation}
The escape probability per unit time of a sticker in a well reads then
\begin{equation} \label{omega}
\Omega(\s)={1\over p}\int_{|{\bf a}|=a} \lambda_{\rm out}(\s+{\bf a}) d^2{\bf a}={t_{\rm out}J(\s)\over I(\s)}
={t_{\rm out}\over a}{u\over \lang(u)}
\end{equation}
where $u=a|\fs|/\kt$, and $\lang(u)=\coth u-u^{-1}$ is the Langevin function. One has $u/ \lang(u)=3+u^2/5+...$ for $u\rightarrow 0$ and $u/ \lang(u)=u+1+o(1)$ for $u\rightarrow \infty$ ; and $u/ \lang(u)\cong(1+2( 1+u^2/4)^{1\over2})$ over the whole range of $u$ to within 3\% precision.

\section*{distribution of strands and  wells}
We model the network as a system of non-interacting strands in a statistically homogeneous distribution of attracting wells. The wells number density $n$ is such that many wells enter the volume spanned in the sticker's Brownian motion. The results of the preceding section need but little modification: using the identity $\int I(\s)d^3\s={4\over3}\pi a^3I_0$, the normalisation integral can be rewritten
\begin{equation}
\widetilde K =I_0+(e^{-w\over\kt}-1)n\int I(\s)d^3\s =(1-x+xe^{-w\over\kt})I_0
\end{equation}
where $x={4\over3}\pi n a^3$ is the volume fraction occupied by the wells. The distribution  for a given strand becomes
\begin{equation}
\phi(\h)=\left\{ \matrix{
{1\over\widetilde K}e^{-{\zo\over\kt}} \qquad\qquad {\rm if}\;\h\;\hbox{\rm outside a well (probability }1-x)
\cr
{1\over\widetilde K}e^{-w\over\kt}e^{-{\zo\over\kt}} \qquad {\rm if}\;\h\;\hbox{\rm inside a well  (probability }x)\ \qquad
\cr
}\right .
\end{equation}
The ensemble of strands is thus described by the connected strand distribution
\begin{equation} \label{psic}
\psic={x\over\widetilde K}e^{-w\over\kt}e^{-{\zo\over\kt}}
\end{equation}
and the free strand distribution
\begin{equation} \label{psif}
\psif={1-x\over\widetilde K}e^{-{\zo\over\kt}}
\end{equation}
satisfying the one-strand normalisation of eq. (\ref{norm}), $\int(\psic+\psif)d^3\h=1$. The ratio
\begin{equation}
{\psic\over\psif}={x\over1-x}e^{-w\over\kt}
\end{equation}
takes the form (\ref{detbal}) upon introducing the connection chemical potential
\begin{equation}
\dz=w+\kt \ln{1-x\over x}
\end{equation}
as the sum of the energy $w$ and the distribution entropy $\kt\ln{1-x\over x}$ accounting for the different volume fraction offered to the connected and to the free stickers. The present model thus satisfies the equilibrium properties announced in the introduction. 

The disconnection rate $\betaf$ is the escape probability per unit time for a connected strand, i.e. a strand whose sticker lies in a well centered on $\s$ such that $|\h-\s|<a$. At lowest order in $a$, we simply have $\Omega(\s)=\Omega(\h)$, and 
\begin{equation}
\betaf={1\over x}\Omega(\h) 
\end{equation}
where the fraction $1/x$ compensates for the normalisation factor $x$ appearing in eq. (\ref{psic}), so that $\betaf\psic=\Omega(\h)\phi(\h\hbox{ in a well})$.
The conection rate $\betac$ relates the probability per unit time of falling into the well to the free strand distribution, so one has similarly
\begin{equation}
\betac={e^{-{w\over\kt}}\over 1-x}\Omega(\h) 
\end{equation}
It must be noted that our $\betaf$ has a much weaker dependence in $\h$ than the exponential form proposed by Tanaka and Edwards (1992b), whereas their $\betac$ is a constant, resulting in a connected strand distribution (their eq. 2.23) incompatible with the Gibbs distribution. 

Our $\betaf$ and $\betac$ are found to satisfy the detailed balance (\ref{detbal}):
\begin{equation} \label{detbal2}
{\betac\over\betaf}={x\over1-x}e^{-w\over\kt}
\end{equation}
In fact, relation(\ref{detbal}) expresses the detailed balance at the global level of connection/disconnection reaction, whereas relation (\ref{detbal2}) is a consequence of the detailed balance at the local level of crossing the barrier at the well border.

\section*{references}
\noindent
Palierne J.F. (2000) Rheothermodynamics of transient networks. submitted to {\it Rheol. Acta=} (2000). posted at {\tt http://xxx.lanl.gov/abs/cond-mat/0007368}
\\ \\
\noindent
Tanaka, F. and Edwards S.F.\\ 
(1992a) Viscoelastic Properties of Physically Cross-Linked Networks. Transient Network Theory. Macromolecules 25: 1516-1523\\
(1992b) Viscoelastic properties of physically cross-linked networks Part 1. Non-linear viscoelasticity. J. Non-Newtonian Fluid Mech. 43: 247-271\\
(1992c) Viscoelastic properties of physically cross-linked networks Part 2. Dynamic mechanical moduli. J. Non-Newtonian Fluid Mech. 43: 273-288\\
(1992d) Viscoelastic properties of physically cross-linked networks Part 3. Time-dependent phenomena. J. Non-Newtonian Fluid Mech. 43: 289-309
\\ \\
\noindent
Yamamoto M. J.\\
(1956) The Visco-elastic Properties of Network Structure I. General formalism. {\it J. Phys. Soc. Jpn} {\bf 11}, 413-421, \\
(1957) The Visco-elastic Properties of Network Structure II. Structural Viscosity. {\it J. Phys. Soc. Jpn} {\bf 12}, 1148-1158, \\
(1958) The Visco-elastic Properties of Network Structure III. Normal Stress Effects (Weissenberg Effect). {\it J. Phys. Soc. Jpn} {\bf 13}, 1200-1211

\end{document}